\documentclass[12pt]{article}
\usepackage{amsmath,amsfonts, feynmp, epsf}
\usepackage{amssymb}
\usepackage{graphicx}
\usepackage{grffile}
\usepackage{array}
\usepackage{makecell}
\newcolumntype{x}[1]{>{\centering\arraybackslash}p{#1}}
\input epsf

\def\bC {\mathbb{C}}

\def\be{\begin{equation}}
\def\ee{\end{equation}}
\def\bea{\begin{eqnarray}}
\def\eea{\end{eqnarray}}
\def\ie{\begin{equation}\begin{aligned}}
\def\fe{\end{aligned}\end{equation}}

\newcommand{\A}{{\alpha}}
\newcommand{\B}{{\beta}}

\newcommand{\da}{{\dot\alpha}}
\newcommand{\db}{{\dot\beta}}

\makeatletter\@addtoreset{equation}{section}\makeatother

\newcommand{\tr}{{\rm tr\,}}

\newcommand{\cG}{{\mathcal G}}

\newcommand{\cZ}{{\mathcal Z}}

\renewcommand{\title}[1]{\vbox{\center\LARGE{#1}}\vspace{5mm}}
\renewcommand{\author}[1]{\vbox{\center#1}\vspace{5mm}}
\newcommand{\address}[1]{\vbox{\center\em#1}}
\newcommand{\email}[1]{\vbox{\center\tt#1}\vspace{5mm}}

\makeatletter\@addtoreset{equation}{section}\makeatother

\begin{document}

\begin{titlepage}
\begin{center}
\hfill \\
\hfill \\
\vskip 1cm

\title{$1/16$ BPS States in ${\cal N}=4$ SYM}

\author{Chi-Ming Chang and
Xi Yin}

\address{Center for the Fundamental Laws of Nature
\\
Jefferson Physical Laboratory, Harvard University,\\
Cambridge, MA 02138 USA}

\email{cmchang@physics.harvard.edu,
xiyin@fas.harvard.edu}

\end{center}

\abstract{ We investigate the problem of counting $1/16$ BPS operators in ${\cal N}=4$ Super-Yang-Mills theory at weak coupling. We present the complete set of $1/16$ BPS operators in the infinite $N$ limit, which agrees with the counting of free BPS multi-graviton states in the gravity dual $AdS_5\times S^5$. Further, we conjecture that all $1/16$ BPS operators in ${\cal N}=4$ SYM are of the multi-graviton form, and give numerical evidences for this conjecture. We discuss the implication of our conjecture and the seeming failure in reproducing the entropy of large $1/16$ BPS black holes in $AdS_5$.
}

\vfill

\end{titlepage}

\section{Introduction}

Holographic dualities \cite{Maldacena:1997re, Gubser:1998bc, Witten:1998qj, Aharony:1999ti} in principle allows for a precise understanding of the microstates of black holes using large $N$ gauge theories. Despite the success in reproducing the Bekenstein-Hawking entropy of certain supersymmetric black holes \cite{Strominger:1996sh}, and various generalizations and refinements, the precise understanding of black hole entropy is largely limited to index computations that are insensitive to the dynamics of the strongly coupled dual gauge theories. It has been suggested that the $1/16$ BPS black holes in $AdS_5\times S^5$ \cite{Gutowski:2004ez, Kunduri:2006ek, Chong:2005da, Chong:2005hr} are of a much richer type: while the states of such black holes should correspond to $1/16$ BPS gauge invariant operators in the dual ${\cal N}=4$ super-Yang-Mills theory at large $N$, the superconformal index that counts the $1/16$ BPS states with signs does not come anywhere near the Bekenstein-Hawking entropy of the black holes. The $1/16$ BPS operators can be counted via the cohomology of one of the supercharges, $Q$, in the weakly coupled gauge theory. The number of BPS operators with a given set of global charges is an integer, and there appears to be no reason why such numbers would jump as one increases the coupling constant. Thus, one would expect the counting of $1/16$ BPS operators at weak coupling to give the same answer as at strong coupling, and reproduces the entropy of the dual black hole. Such a counting nonetheless depends crucially on the particular form of the interactions in the gauge theory, as the weak coupling answer is entirely different from the free field theory answer, and one could hope to learn about the structure of multi-trace operators responsible for the entropy of the black hole. For earlier attempts see \cite{Berkooz:2006wc, Kim:2006he, Grant:2008sk}.

In this paper, we report on a renewed attempt at this counting problem purely in the gauge theory, closely following the approach of \cite{Grant:2008sk}. Firstly, we reformulate the cohomology of the supercharge $Q$ in terms of a relative Lie algebra cohomology that involves the infinite dimensional Lie superalgebra $\cG_N=\bC[z_+,z_-]\otimes \Lambda[\theta_1,\theta_2,\theta_3]\otimes sl_N$, and we derive a complete set of $Q$-cohomology classes in the infinite $N$ limit, and match them with the multi-graviton states in the bulk. While this agreement is widely expected, to the best of our knowledge no complete derivation previously existed in the literature (for earlier work see \cite{Janik:2007pm}). However, we have not been able to find, nor have seen any evidence for, ``new" $Q$-cohomology classes that are not of the multi-graviton form. Our formulation of the counting problem allows for straightforward (but extremely time consuming) computer tests. In all examples of low dimension operators in $SU(2)$, $SU(3)$, or $SU(4)$ gauge theories we have tested, no new cohomology is found. We are thus led to conjecture that the complete spectrum of $1/16$ BPS operators in ${\cal N}=4$ SYM are of the multi-graviton form. 

If our conjecture is correct, the number of $1/16$ BPS states at weak coupling is much less than what is needed to account for the entropy of the dual black holes. The failure to produce the black hole entropy, or even the correct scaling with $N$, appears to be a puzzle. We comment on possible resolutions at the end of the paper.

\section{Organizing the $Q$ action on letters}

We consider $SU(N)$ ${\cal N}=4$ super-Yang-Mills theory, with 16 supercharges $Q^i_\A$, $\overline Q_{i\da}$, and 16 special supercharges $S^\A_i, \overline{S}^{i\da}$. The index $i$ runs from 1 to 4, and we will write the spinor index as $\A=\pm$. Following the notation of \cite{Grant:2008sk}, we will be considering $1/16$ BPS operators (or states in radial quantization) that are annihilated by $Q\equiv Q^4_-$ and $S \equiv S_4^-$. In radial quantization, $S=Q^\dagger$, and
\ie
2\{Q, Q^\dagger\} = \Delta \equiv E - 2J - H_1 - H_2 - H_3,
\fe
where $E$ is the conformal weight of the operator, $J=J_L^3$ is the left $SU(2)_L$ angular momentum, and $H_1, H_2, H_3$ are the Cartan generators of the $SO(6)$ R-symmetry. The $1/16$ BPS operators are in one-to-one correspondence with $Q$-cohomology classes on the set of gauge invariant operators.

The fields of ${\cal N}=4$ SYM consists of 6 scalars $\Phi_{ij}$, obeying the reality condition $\Phi^{ij}\equiv \Phi_{ij}^* = {1\over 2}\epsilon^{ijkl} \Phi_{kl}$, 4 chiral fermions $\Psi_{i\A}$, their complex conjugates $\bar \Psi^{i\da}$, and the gauge field $A_{\A\db}$. In the weak coupling limit, we only need to consider operators that are made out of BPS ``letters", namely the component fields and gauge covariant derivatives whose classical dimension and charges saturate the BPS bound $\Delta=0$. These are
\ie
\phi^n\equiv \Phi^{4m},~~~~ \psi_n\equiv -i\Psi_{n+}, ~~~n=1,2,3,~~~~ \lambda_\da = \bar\Psi^4_{\da},~~~~ f\equiv -iF_{++},
\fe
along with the covariant derivatives
\ie
D_\da \equiv D_{+\da}.
\fe
These letters are subject to the relations
\ie{}
[D_\da, D_\db] = \epsilon_{\da\db} f,~~~~ D_\da \lambda^\da = [\phi^n, \psi_n],
\fe
where the second equation is the only equation of motion that is purely made out of BPS letters. The action of the supercharge $Q$ on the relevant component fields are given by
\ie
&[Q,\phi^n]=0,~~~~\{Q,\psi_n\}=-i\epsilon_{nmp}[\phi^m,\phi^p],
\\
&\{Q,\lambda_\da\}=0,~~~~[Q,f]=i[\phi^n,\psi_n].
\fe
The action of $Q$ on the covariant derivative is given by
\ie
&[Q,D_\da \zeta]=-i[\lambda_\da,\zeta]+D_\da Q \zeta.
\fe

Since we can trade the commutator of covariant derivatives with a field strength, it suffices to consider symmetrized covariant derivatives acting on the fields. Likewise, when $D_\da$ acts on $\lambda_\db$, the part that is antisymmetric in $\da\db$ can be traded with the commutator of $\phi^n$ with $\psi_n$ using the equation of motion, and thus it suffices to consider the covariant derivatives of $\lambda_\db$ where the spinor index $\db$ are completely symmetrized with all the spinor derivatives of the derivatives. Now, to organize the $Q$-action on these fields as well as their derivatives, we introduce auxiliary commuting variables $z^\da$, and the generating fields
\ie
&\phi^m(z)=\sum^\infty_{n=0}{1\over n!}(z^\da D_\da)^n\phi^m,~~~\psi_m(z)=\sum^\infty_{n=0}{1\over n!}(z^\da D_\da)^n\psi_m,
\\
&\lambda(z)=\sum^\infty_{n=0}{1\over (n+1)!}(z^\da D_\da)^n(z^\db\lambda_\db),~~~f(z)=\sum^\infty_{n=0}{1\over n!}(z^\da D_\da)^nf.
\fe
Now all independent gauge invariant operators can be obtained by taking the product of traces of products of $z$-derivatives of these generating fields. The generating fields are unconstrained, except for the condition $\lambda(0)=0$.

The $Q$ action on these generating fields are given by
\ie
&[Q,\phi^m(z)]=-i[\lambda(z),\phi^m(z)],
\\
&[Q,f(z)]=-i[\lambda(z),f(z)]+i[\phi_n(z),\psi_n(z)],
\\
&\{Q,\psi_m(z)\}=-i\{\lambda(z),\psi_m(z)\}-i\epsilon_{mnp}[\phi_n(z),\phi_p(z)],
\\
&\{Q,\lambda(z)\}=-i\lambda(z)^2.
\fe

The $z$-dependent generating fields $\phi^n(z), \psi_n(z), \lambda(z), f(z)$ can be further organized into a single generating $(2|3)$ ``superfield" $\Psi(z,\theta)$, as
\ie
\Psi(z,\theta)=-i\left[\lambda(z)+2\theta_n\phi^n(z)+\epsilon^{mnp}\theta_m\theta_n\psi_p(z)+4\theta_1\theta_2\theta_3 f(z)\right].
\fe
Here $\theta_1, \theta_2, \theta_3$ are three anti-commuting variables. We will write collectively ${\cal Z}=(z,\theta)$.
One can verify that the action of $Q$ on the superfield $\Psi({\cal Z})$ takes an extremely concise form:
\ie\label{Qaction}
\{Q,\Psi(\cZ)\}=\Psi(\cZ)^2.
\fe
The only constraining condition on $\Psi({\cal Z})$ is
\ie\label{CNS}
\left.\Psi\right|_{z^\da=\theta_n=0}=0.
\fe
Now all gauge invariant operators can be built from products of $z$ or $\theta$ derivatives of $\Psi({\cal Z})$, and then setting ${\cal Z}=0$.

\section{BPS states as Lie algebra cohomology}
The $Q$-cohomology on the space of words constructed out of BPS letters can be rephrased in the language of Lie algebra cohomology. Let $\cG$ be a Lie algebra and $M$ be a $\cG$-module. A $M$-valued $p$-cochain on $\cG$ is a skew-symmetric $p$-linear map
\ie
c:\cG\wedge\shortstack[pos]{$p$\\ $\cdots$}\wedge\cG\to M.
\fe
The (abelian) group of all $p$-cochains is denoted by ${\rm C}^p(\cG;M)$, i.e.
\ie
{\rm C}^p(\cG;M)= \text{Hom}(\Lambda^p\cG,M).
\fe
The Lie algebra cohomology ${\rm H}^p(\cG,M)$ on $\cG$ is defined by the cohomology on the complex ${\rm C}^*(\cG;M)$ with the differential $d$, where for $c\in {\rm C}^p(\cG;M)$, $dc\in {\rm C}^{p+1}(\cG;M)$ is defined by
\ie\label{Cartan}
dc(x_1,\cdots,x_{p+1})=&\sum_{1\le i<j\le p+1}(-1)^{i+j}c([x_i,x_j],x_1,\cdots,\hat x_i,\cdots,\hat x_j,\cdots,x_{p+1})
\\
&+\sum^{p+1}_{i=1}(-1)^{i+1}x_i\cdot c(x_1,\cdots,\hat x_i,\cdots,x_{p+1}).
\fe
Let ${\cal H}$ be a subalgebra of $\cG$. A relative (to the subalgebra ${\cal H}$) $p$-cochain is a $p$-cochain $c\in {\rm C}^p(\cG;M)$ satisfying the conditions
\ie\label{RLAC1}
&c(g_1,\cdots,g_{p-1},h)=0
\fe
if $h\in {\cal H}$, and
\ie\label{RLAC2}
&c([h,g_1],g_2,\cdots,g_p)+\cdots+c(g_1,\cdots,g_{p-1},[h,g_p])-h\cdot c(g_1,\cdots,g_p)=0
\fe
for all $h\in {\cal H}$. The group of all relative $p$-cochains is denoted by ${\rm C}^p(\cG,{\cal H};M)$. The relative Lie algebra cohomology is defined as the cohomology on the relative complex ${\rm C}^*(\cG,{\cal H};M)$ with the differential $d$ defined on \eqref{Cartan}.

In the language of Lie algebra cohomology, the $(2|3)$ superfield $\Psi$ is an element of a (infinite dimensional) Lie algebra $\cG_N\equiv\bC[z_+,z_-]\otimes \Lambda[\theta_1,\theta_2,\theta_3]\otimes sl_N$. The words constructed out of the BPS letters are the cochains in the space ${\rm C}^p(\cG_N;\bC)$. The $Q$-action \eqref{Qaction} on $\Psi$ gives the differential \eqref{Cartan} on the words. The constraint \eqref{CNS} and the requirement of gauge invariance on the words are exactly the two conditions on \eqref{RLAC1} and \eqref{RLAC2} with the subalgebra being $sl_N$. Notice that the terms on the second line of \eqref{Cartan} and the last term of \eqref{RLAC2} vanish due to the triviality of the coefficient. Hence, our $Q$-cohomology classes are given by the relative Lie algebra cohomology
\ie
{\rm H}^*(\cG_N,sl_N;\bC).
\fe
The degree of the cohomology class is given by the degree of the corresponding BPS operator in $\Psi$.

\section{$Q$ cohomology at infinite $N$}

The energy $E$, $SU(2)_L$ and $SU(2)_R$ angular momenta $J_L^3$, $J_R^3$, and the charges with respect to the three $SO(6)_R$ Cartan generators, $q_1, q_2, q_3$, of the relevant letters are listed in the following table.

\centerline{\begin{tabular}{|x{1cm}|x{1cm}|x{1cm}|x{1cm}|x{1.5cm}|}\hline
 & $E$ & $J_L^3$ & $J_R^3$ & $q_i$ 
 \\ \hline
 $\lambda_\pm$ & ${3\over 2}$ & $0$ & $\pm{1\over 2}$ & ${1\over 2}$
\\ \hline 
 $\phi^n$ & $1$ & $0$ & $0$ & $\delta_{in}$
\\ \hline
 $\psi_n$ & $3\over 2$ & ${1\over 2}$ & $0$ & ${1\over 2}-\delta_{in}$
\\ \hline
 $f$ & $2$ & $1$ & $0$ & $0$
\\ \hline
 $D_\pm$ & $1$ & ${1\over 2}$ & $\pm {1\over 2}$ & $0$
\\ \hline
\end{tabular}
}
\bigskip
The charge assignments on $Q$, the generating superfield $\Psi$, and the generating parameters $z, \theta$ are listed as follows.
\bigskip

\centerline{\begin{tabular}{|x{1cm}|x{1cm}|x{1cm}|x{1cm}|x{1.5cm}|}\hline
 & $E$ & $J_L^3$ & $J_R^3$ & $q_i$
 \\ \hline
 $Q$ & $1\over 2$ & $-{1\over 2}$ & 0 & ${1\over 2}$
 \\ \hline
 $\Psi$ & $1\over 2$ & $-{1\over 2}$ & 0 & ${1\over 2}$
\\ \hline 
 $z_\pm$ & $-1$ & $-{1\over 2}$ & $\pm{1\over 2}$ & $0$
\\ \hline
 $\theta_n$ & $-{1\over 2}$ & $-{1\over 2}$& 0  & ${1\over 2}-\delta_{in}$
\\ \hline
\end{tabular}
}
\bigskip

Within our sector that consists of letters that saturate the BPS bound in the free limit, there is an extra ``bonus" $U(1)$ symmetry (see \cite{Intriligator:1998ig, Intriligator:1999ff} for related discussions), that assigns charge 1 to $Q$ and $\Psi$, and charge 0 to $z_\da$ and $\theta_n$. We will denote this charge by $Y$.

Let us define the following ``refined" partition function of $1/16$ BPS states, that takes into account the bonus charge $Y$,
\ie
Z(x,a,b,u,v,w)={\rm Tr}_{{1\over16}{\rm-BPS}}\left[
x^Y a^{E-J_L^3+J_R^3-Y} b^{E-J_L^3-J_R^3-Y} u^{-q_2-q_3+Y} v^{-q_1-q_3+Y} w^{-q_1-q_2+Y}
\right].
\fe
In terms of the generating superfield $\Psi({\cal Z})$, the power of $x$ simply counts the number of $\Psi$'s in the operator, $a,b$ count the number of $z$-derivatives $\partial_{z_+}$, $\partial_{z_-}$, while $u,v,w$ count the number of $\theta$-derivatives, $\partial_{\theta_1},\partial_{\theta_2},\partial_{\theta_3}$. 

Note that $Q$ simply increases the degree in $x$. An index can thus be defined by restricting the partition function to the special case $x=-1$. In this case, the trace over only the $1/16$-BPS states is equal to the trace over all states, and we can define our refined index to be
\ie
&{\cal I}(a,b,u,v,w) = Z(-1,a,b,-u,-v,-w)
\\
&={\rm Tr}\left[
(-1)^{F} a^{E-J_L^3+J_R^3-Y} b^{E-J_L^3-J_R^3-Y} u^{-q_2-q_3+Y} v^{-q_1-q_3+Y} w^{-q_1-q_2+Y}
\right].
\fe
The superconformal index of ${\cal N}=4$ SYM defined in \cite{Kinney:2005ej} is related by
\ie
{\cal I}_{YM}(t,y,v,w) = {\rm Tr}\left[ (-1)^F t^{2(E+J_L^3)} y^{2J_R^3} v^{R_2} w^{R_3} \right] =  {\cal I}(t^3y,t^3/y,t^2/w,t^2w/v,t^2v).
\fe

To find $Q$-cohomology classes, let us inspect the $Q$ action on $\theta_n$ and $z_\da$ derivatives of $\Psi$:
\ie\label{QactiononPPsi}
&[Q,\partial_{\theta_n}\Psi(\cZ)]=[\Psi(\cZ),\partial_{\theta_n}\Psi(\cZ)],
\\
&\{Q,\partial_{z^\A}\Psi(\cZ)\}=\{\Psi(\cZ),\partial_{z^\A}\Psi(\cZ)\},
\\
&\{Q,\partial_{\theta_m}\partial_{\theta_n}\Psi(\cZ)\}=\{\Psi(\cZ),\partial_{\theta_m}\partial_{\theta_n}\Psi(\cZ)\}-[\partial_{\theta_m}\Psi(\cZ),\partial_{\theta_n}\Psi(\cZ)],
\\
&\{Q,\partial_{z^\A}\partial_{z^\B}\Psi(\cZ)\}=\{\Psi(\cZ),\partial_{z^\A}\partial_{z^\B}\Psi(\cZ)\}+\{\partial_{z^\A}\Psi(\cZ),\partial_{z^\B}\Psi(\cZ)\},
\\
&[Q,\partial_{z^\A}\partial_{\theta_n}\Psi(\cZ)]=[\Psi(\cZ),\partial_{z^\A}\partial_{\theta_n}\Psi(\cZ)]+[\partial_{z^\A}\Psi(\cZ),\partial_{\theta_n}\Psi(\cZ)].
\fe
A class of nontrivial single-trace $Q$-closed operators are
\ie\label{stq}
(\partial_{z_+})^{p_1}(\partial_{z_-})^{p_2}(\partial_{\theta_1})^{q_1}(\partial_{\theta_2})^{q_2}(\partial_{\theta_3})^{q_3} {\rm Tr}\left[(\partial_{z_+}\Psi)^{k_1}(\partial_{z_-}\Psi)^{k_2}\left(\partial_{\theta_1}\Psi\right)^{m_1}\left(\partial_{\theta_2}\Psi\right)^{m_2}\left(\partial_{\theta_3}\Psi\right)^{m_3}\right]\Big|_{z=\theta=0},
\fe
where $q_1, q_2, q_3$ and $k_1, k_2$ are restricted to be $0$ or $1$. By \eqref{QactiononPPsi}, if we change the order of the $\partial\Psi$'s inside the trace, the difference is given by a $Q$-exact operator. Therefore, a nontrivial $Q$-cohomology representative is given by the operator of the form (\ref{stq}) with the $\partial_{\cZ}\Psi$'s symmetrized or anti-symmetrized inside the trace according to their statistics.

In the infinite $N$ limit, where we can ignore trace relations, (\ref{stq}) gives a complete set of representatives of $Q$-cohomology on single-trace operators. The general $Q$-cohomology elements are then given by products of single trace expressions of the form (\ref{stq}). The partition function of all single-trace operators of the form (\ref{stq}) is
\ie\label{zst}
&Z_{st}(x,a,b,u,v,w)={(1+u)(1+v)(1+w)\over (1-a)(1-b)}\left[{(1+ax)(1+bx)\over (1-ux)(1-vx)(1-wx)}-1\right]
\\
&~~~~ -{u(1+v)(1+w)\over (1-a)(1-b)}\left[{(1+ax)(1+bx)\over (1-ux)(1-vx)(1-wx)}-1\right]
\\
&~~~~ -{v(1+w)\over (1-a)(1-b)}\left[{(1+ ax)(1+ bx)\over (1-vx)(1-wx)}-1\right]
\\
&~~~~ -{w\over (1-a)(1-b)}\left[{(1+ ax)(1+ bx)\over 1- wx}-1\right]-{ab\over(1-a)(1-b)}x.
\fe
The counting goes as follows. The first term on the RHS of (\ref{zst}) counts all the traces of a string of $\partial\Psi$'s, with all 5 types of derivatives $\partial_{z_\da}, \partial_{\theta_n}$ acting outside the trace, while ignoring relations among the various derivatives of trace of derivatives. The second term on the RHS of (\ref{zst}) subtracts off the contribution from operators involving a $\partial_{\theta_1}$ acting outside the trace, because these would be double counting, due to the relation
\ie\label{rel}
0=&k_1\partial_{z^1}\tr\left[(\partial_{z^1}\Psi)^{k_1-1}(\partial_{z^2}\Psi)^{k_2}\left(\partial_{\theta_1}\Psi\right)^{m_1}\left(\partial_{\theta_2}\Psi\right)^{m_2}\left(\partial_{\theta_3}\Psi\right)^{m_3}\right]
\\
&-k_2\partial_{z^2}\tr\left[(\partial_{z^1}\Psi)^{k_1}(\partial_{z^2}\Psi)^{k_2-1}\left(\partial_{\theta_1}\Psi\right)^{m_1}\left(\partial_{\theta_2}\Psi\right)^{m_2}\left(\partial_{\theta_3}\Psi\right)^{m_3}\right]
\\
&+m_1\partial_{\theta_1}\tr\left[(\partial_{z^1}\Psi)^{k_1}(\partial_{z^2}\Psi)^{k_2}\left(\partial_{\theta_1}\Psi\right)^{m_1-1}\left(\partial_{\theta_2}\Psi\right)^{m_2}\left(\partial_{\theta_3}\Psi\right)^{m_3}\right]
\\
&+m_2\partial_{\theta_2}\tr\left[(\partial_{z^1}\Psi)^{k_1}(\partial_{z^2}\Psi)^{k_2}\left(\partial_{\theta_1}\Psi\right)^{m_1}\left(\partial_{\theta_2}\Psi\right)^{m_2-1}\left(\partial_{\theta_3}\Psi\right)^{m_3}\right]
\\
&+m_3\partial_{\theta_3}\tr\left[(\partial_{z^1}\Psi)^{k_1}(\partial_{z^2}\Psi)^{k_2}\left(\partial_{\theta_1}\Psi\right)^{m_1}\left(\partial_{\theta_2}\Psi\right)^{m_2}\left(\partial_{\theta_3}\Psi\right)^{m_3-1}\right],
\fe
where the traces are understood to be completely (anti-)symmetrized.
Similarly, the third term on the RHS of (\ref{zst}) eliminates the double counting due to $\partial_{\theta_2}$ acting on a trace that does not involve $\partial_{\theta_1}\Psi$; the fourth term eliminates $\partial_{\theta_3}$ acting on a trace that does not involve $\partial_{\theta_1}\Psi$, $\partial_{\theta_2}\Psi$; finally, the last term in (\ref{zst}) eliminates the double counting due to the relation $\partial_{z_1}{\rm Tr}(\partial_{z_2}\Psi) = \partial_{z_2}{\rm Tr}(\partial_{z_1}\Psi)$.
Though not immediately obvious, the formula (\ref{zst}) is indeed symmetric in $(u,v,w)$ and in $(a,b)$.

(\ref{zst}) indeed agrees with the partition function of a single $1/16$ BPS graviton in $AdS_5\times S^5$; namely, the single particle partition function $Z_{sp}$ (in (5.13) of \cite{Kinney:2005ej}) agrees with the single-trace partition function:
\ie
Z_{sp}=Z_{st}(x/z,x^2yz,x^2z/y,xz/w,xzw/v,xzv).
\fe
In particular, the ``blind" partition function is
\ie
&{\rm Tr}_{st}(x^{2E}) = Z_{st}(x,x^2,x^2,x,x,x)
\\
&={(1+x)^3\over (1-x^2)^2}\left[{(1+x^3)^2\over (1-x^2)^3}-1\right]-{x(1+x)^2\over (1-x^2)^2}\left[{(1+x^3)^2\over (1-x^2)^3}-1\right]-{x(1+x)\over (1-x^2)^2}\left[{(1+x^3)^2\over (1-x^2)^2}-1\right]
\\
&~~~~ -{x\over (1-x^2)^2}\left[{(1+x^3)^2\over 1-x^2}-1\right]-{x^2\over (1-x^2)^2}x^3
\\
&= {x^2(3-2x^2+8x^4-2x^6+x^8) + x^3(5-2x^2+5x^4)\over (1-x)^5}.
\fe
Therefore, we have reproduced the correct counting of $1/16$ BPS gravitons in $AdS_5\times S^5$ from the $Q$ cohomology in the infinite $N$ limit.

\section{The conjecture at finite $N$ and checks}

At finite $N$, a class of representatives of $Q$-cohomology classes are given by the ``multi-graviton" operators, of the form
\ie\label{mg}
\prod_i (\partial_{z_+})^{p_1^{(i)}}(\partial_{z_-})^{p_2^{(i)}}(\partial_{\theta_1})^{q_1^{(i)}}(\partial_{\theta_2})^{q_2^{(i)}}(\partial_{\theta_3})^{q_3^{(i)}} {\rm Tr}\left[(\partial_{z_+}\Psi)^{k_1^{(i)}}(\partial_{z_-}\Psi)^{k_2^{(i)}}\left(\partial_{\theta_1}\Psi\right)^{m_1^{(i)}}\left(\partial_{\theta_2}\Psi\right)^{m_2^{(i)}}\left(\partial_{\theta_3}\Psi\right)^{m_3^{(i)}}\right]\Big|_{z=\theta=0}.
\fe
There are generally relations among these operators, up to $Q$-exact terms, at finite $N$. Such relations eliminates some of the multi-graviton states. The question is, are there any new $Q$ cohomology that are not of the multi-graviton form, at finite $N$? If the counting $1/16$ BPS states at weak coupling are to match with the Bekenstein-Hawking entropy of the known large $1/16$ BPS black hole solutions, it appears that most of the states at dimension $E \sim N^2$ would have to come from the new $Q$-cohomology, since the entropy of a gas of gravitons scales like $E^{5/6}$ which is much less than the Bekenstein-Hawking entropy of the black hole that goes like $N^2$.

If there are such new $Q$-cohomology, they must come from operators that are $Q$-closed only due to trace relations, and are themselves not a trace relation up to $Q$-exact terms. More precisely, let us consider multi-trace expressions in the $U(N)$ or $SU(N)$ theory, and decrease $N$ by 1 at a time.


Let $V_N$ be the module of all multi-trace operators constructed out of $\partial_\cZ^n \Psi|_{\cZ=0}$ in the $SU(N)$ theory, and $V_{N-1}$ for the $SU(N-1)$ theory. The obvious embedding $SU(N-1)\subset SU(N)$ gives rise to a reduction $\pi:V_N \to V_{N-1}$ due to new trace relations of $SU(N-1)$.
Let $R_N$ be the ideal of such new trace relations, when $\Psi$ is reduced from an $N\times N$ to an $(N-1)\times(N-1)$ traceless matrix. In other words, we have a short exact sequence
\ie\label{ses}
0\to R_N\to V_N\stackrel{\!\pi}{\rightarrow} V_{N-1}\to 0.
\fe
It is not hard to see that $Q$ commutes with $\pi$ as well as the inclusion $R_N\to V_N$. (\ref{ses}) then induces a long exact sequence on the $Q$-cohomology, graded say by the degree in $\Psi$,
\ie
\cdots\to {\rm H}^n_Q(R_N)\to {\rm H}^n_Q(V_N)\to {\rm H}^n_Q(V_{N-1})\stackrel{\!f}{\rightarrow} {\rm H}^{n+1}_Q(R_N)\to\cdots.
\fe
The existence of new $Q$ cohomology is equivalent to the map
\ie
f: {\rm H}_Q^n(V_{N-1}) \to {\rm H}_Q^{n_1}(R_N)
\fe
being nonzero.

If there exists an adjoint operator $Q^{ad}$ of $Q$ with respect to some inner product on $V_N$, that commutes with $\pi$, then $f$ must be zero. Since ${\rm H}^*_Q(V_N)$ is isomorphic  to $\text{ker}\{Q,Q^{ad}\}$, the map between ${\rm H}^*_Q(R_N)$ and ${\rm H}^*_Q(V_N)$ would become an inclusion $\text{ker}\{Q,Q^{ad}\}\big|_{R_N}\hookrightarrow
\text{ker}\{Q,Q^{ad}\}$ if $Q^{ad}$ commutes with $\pi$. Note that $Q$ can be schematically written in the form
\ie
{\rm Tr}\left(\Psi\Psi{\partial\over \partial \Psi}\right).
\fe
One may try to work with the inner product treating each $\Psi$ (more precisely, each $\partial_\cZ^n\Psi|_{\cZ=0}$) as a creation operator on a Fock space, and the adjoint operator
\ie
Q^{ad}={\rm Tr}\left(\Psi{\partial\over \partial \Psi}{\partial\over \partial \Psi}\right).
\fe
However, such a $Q^{ad}$ does not commute with $\pi$, and thus such a naive attempt at proving the non-existence of new $Q$ cohomology fails.

Nonetheless, we have not been able to find any new cohomology, by enumerating low dimension operators in the $SU(2), SU(3)$ (and a few examples in $SU(4)$) cases, neither do we see any evidence for the existence of new $Q$ cohomology. The enumeration of cohomology classes in $SU(2)$ and $SU(3)$ examples are summarized in Appendix A.

As an example of failure in finding new $Q$ cohomology, consider the $SU(N)$ theory. A fermionic matrix $X$ is subject to the trace relation that ${\rm Tr} (X^{2N+1})$ can be written in terms of multi-trace operators (or simply vanishes). Applying this to ${\rm Tr}((\partial_{z_+}\Psi)^{2N+1}) = Q {\rm Tr}(\partial_{z_+}^2\Psi (\partial_{z_+}\Psi)^{2N-1})$, we may ask if ${\rm Tr}(\partial_{z_+}^2\Psi (\partial_{z_+}\Psi)^{2N-1})$ gives rise to a new $Q$-cohomology class. But in fact, for a pair of fermionic matrices $X$ and $Y$, there is also a trace relation relating ${\rm Tr} (X^{2N-1} Y)$ to multi-trace operators, and no new cohomology arises this way.

We are thus led to conjecture that the multi-graviton operators (\ref{mg}) do in fact give the complete set of $Q$ cohomology at any finite $N$, and that {\it there are no new $Q$ cohomology due to trace relations}.

A test of the conjecture would be to reproduce the refined index ${\cal I}(a,b,u,v,w)$ from the multi-graviton operators, subject to trace constraints. The index is easily computed in our formalism, by considering the words made out of the letters $\partial_\cZ^k\Psi$, 
\ie\label{indd}
{\cal I}(a,b,u,v,w) = \int_{SU(N)} dU \exp\left\{ \sum_{n=1}^n \left[ 1- {(1-u^n)(1-v^n)(1-w^n)\over (1-a^n)(1-b^n)} \right]{ {\rm Tr}(U^n){\rm Tr}(U^{-n})\over n} \right\}.
\fe
This should count with signs the polynomials in $\partial_\cZ^k \eta_i$ ($i=1,2,\cdots,N$) of the form
\ie\label{sympol}
\prod \left[\partial_{z_+}^{p_1}\partial_{z_-}^{p_2} \partial_{\theta_1}^{q_1}\partial_{\theta_2}^{q_2}\partial_{\theta_3}^{q_3} \sum_{i=1}^N (\partial_{z_+} \eta_i)^{k_1}(\partial_{z_-} \eta_i)^{k_2}(\partial_{\theta_1} \eta_i)^{m_1}(\partial_{\theta_2} \eta_i)^{m_2}(\partial_{\theta_3} \eta_i)^{m_3}\right],
\fe
where the odd variables $\eta_i$ may be thought of as eigenvalues of $\Psi$ (subject to the constraint $\eta_1+\eta_2+\cdots +\eta_N=0$ in the case of $SU(N)$ theory), since the (anti-)commutators of $\partial_\cZ\Psi$'s within each trace are ignored. The direct counting of all such polynomials is nontrivial, however, since not all symmetric polynomials in higher order derivatives $\partial_\cZ^k \eta_i$ are of the form (\ref{sympol}). We conjecture that (\ref{indd}) is in fact the answer to this counting problem (when counted with sign).

We can easily check the index in some simple special cases. In the sector that only involves $z_+$ derivatives, only a single $\partial_{z_+}\Psi$ can appear in the trace, and the trace vanishes in $SU(N)$ theory. Indeed, one can verify that
\ie
{\cal I}(a,0,0,0,0)=1.
\fe
Another special case is the sector that only involves $\theta_1$ derivatives. The multi-graviton operators are products of the bosonic operators ${\rm Tr}((\partial_{\theta_1}\Psi)^n)$, and no $\partial_{\theta_1}$ outside the trace. Such operators subject to $SU(N)$ trace relations are straightforward to count with a matrix model, and the answer is indeed
\ie
{\cal I}(0,0,u,0,0) = \int_{SU(N)} dU \exp\left[ \sum_{n=1}^n u^n {{\rm Tr}(U^n){\rm Tr}(U^{-n})\over n} \right].
\fe

We can also put a upper bound on the number of BPS multi-graviton operators of given set of charges using the representation (\ref{sympol}), by counting {\it all} symmetric polynomials in $\partial_{z_+}^{k_1}\partial_{z_-}^{k_2} \partial_{\theta_1}^{m_1}\partial_{\theta_2}^{m_2}\partial_{\theta_3}^{m_3} \eta_i$, $i=1,2,\cdots,N$ (rather than  symmetric polynomials of the restricted form (\ref{sympol})). The symmetric polynomials are counted by the partition function
\ie
&Z_{sym}(x,a,b,u,v,w) 
\\
&= \left.\prod_{n_{(k,m)}} \left[ 1- p \prod_{k_{1,2}\geq 0, ~ m_{1,2,3}=0,1} \left(-x a^{k_1} b^{k_2} (-u)^{m_1} (-v)^{m_2} (-w)^{m_3}\right)^{n_{(k,m)}} \right]^{-(-1)^{\sum n_{(k,m)}(1+m_1+m_2+m_3)}}\right|_{p^N},
\fe
where the range of $n_{(k,m)}\equiv n_{(k_1,k_2;m_1,m_2,m_3)}$ is given by
\ie
& n_{(k,m)}=0,1,2,\cdots,\infty,~~~ m_1+m_2+m_3={\rm odd},
\\
& n_{(k,m)}=0,1,~~~~~~m_1+m_2+m_3={\rm even}.
\fe
The entropy $S$ of these objects in the large charge/energy limit ($E\gg N$) can be estimated using the same thermodynamical approximation as in \cite{Grant:2008sk}, yielding $S\sim N^{1/3} E^{2/3}$ (the same scaling in $N$ and $E$ as in the purely scalar sector, but with a different coefficient). This bound is better than the one given by the entropy of free gravitons ($\sim E^{5/6}$) when $E$ is greater than $N^2$.

\section{Discussion}

Our conjecture that there are no new $Q$ cohomology due to trace relations, which would imply that all $1/16$ BPS states in ${\cal N}=4$ SYM are of the multi-graviton form, is in apparent conflict with the existence of large $1/16$ supersymmetric black holes in $AdS_5\times S^5$ and their Bekenstein-Hawking entropy. The computer test of the conjecture in the $SU(N)$ theory for any given number $N$ and for cohomology classes of any given set of charges is in principle straightforward, but in practice extremely time consuming and costly in terms of memory. One possibility is that new cohomology classes, say for $SU(3)$ theory, only show up at very large charges and evaded our tests so far.

Our conjecture is equivalent to the statement that for the $Q$ operator defined as $Q\Psi(\cZ)=\Psi(\cZ)^2$ on the space of gauge invariant polynomials in the coefficients of the fermionic matrix power series $\Psi(\cZ)$ in $(2|3)$-superspace variable $\cZ$, no new cohomology shows up when one restricts from the $SU(N)$ to the $SU(N-1)$ case. It doesn't seem that the $(2|3)$ superspace is of any particular significance, and our conjecture naturally extends to the case of $(n|m)$ superspace variable $\cZ$, though the $Q$ cohomology in the latter more general case would not be related to $1/16$ BPS states in ${\cal N}=4$ SYM. If our conjecture is correct, then there appear to be two logical possibilities to reconcile with the black hole entropy in the bulk.

One possibility is that there are jumps in the number of $Q$ cohomology of given charges as one moves to strong coupling. Such a phenomenon would be extremely interesting, but we do not see any evidence for it.

Another possibility is that the large $1/16$ BPS black holes exists only in the supergravity limit and not in the full string theory. It could be that $\alpha'$ corrections to the supergravity equations does not allow for the corresponding (deformed) extremal black hole solutions to maintain their supersymmetries. If this is the case, we would expect that while the number of exactly $1/16$ BPS states is small, there are a large number of very near $1/16$ BPS states at strong 't Hooft coupling that account for the entropy of the large charged black hole. This would be somewhat counter intuitive, however, since one expects the anomalous dimension of the near-BPS operators to grow with the 't Hooft coupling. To understand how the spectrum of near $1/16$ BPS operators of dimension $\sim N^2$ changes with the coupling requires a much more detailed understanding of the dynamics of the gauge theory (which, a priori, goes beyond the applicable regime of integrability methods \cite{Beisert:2010jr}).

In summary, while our result is largely a negative one: we had not learned anything new about the structure of high dimension multi-trace operators that correspond to the typical microstates of large black holes in $AdS_5$, our puzzle is sharp and we hope it can be resolved in the near future.

\bigskip

\section*{Acknowledgments}

We are extremely grateful to Shiraz Minwalla for drawing our interest to this problem, and to Juan Maldacena and Shiraz Minwalla for helpful discussions and comments on a preliminary draft. The numerical tests are performed using Mathematica 8.0.1 on the Harvard Odyssey Cluster. This work is supported in part by the Fundamental Laws Initiative Fund at Harvard University.
X.Y. is supported in part by NSF Award PHY-0847457, and by a Sloan Fellowship.

\appendix

\section{Some low level examples}

Here we list some explicit enumeration of $Q$ cohomology in the $SU(N)$ theory of given charges, for $N=2,3$. The first line of each table lists the charge vectors that count the number of derivatives $(\partial_{z_+}, \partial_{z_-}; \partial_{\theta_1}, \partial_{\theta_2}, \partial_{\theta_3})$. In each row below, we list the number of $Q$-cohomology classes in the $SU(N)$ case, of degree $(2,3,4,\dots)$ in $\Psi$. Note that there are no states of degree less than 2 in $\Psi$, and the maximal allowed degree in $\Psi$ is equal to the total number of $(z,\theta)$ derivatives. In all of the examples we have tested, all $Q$-cohomology classes are represented by multi-graviton states, and no new $Q$-cohomology due to trace relations are found. The limitation on carrying out the test to higher levels is entirely due to computer time and memory. Note that within more restricted charge sectors, one could go to higher levels.

\bigskip
\bigskip

\bigskip

\centerline{\begin{tabular}{ |x{2.0cm}|x{3.4cm}|x{3.4cm}|x{3.8cm}|}\hline
charges & $N=2$ & $N=3$ & $N=\infty$
\\ \hline
$[4,4;0,0,0]$ & $(1,0,4,0,0,0,0)$ & $(1,0,5,0,1,0,0)$ & $(1,0,5,0,2,0,1)$ 
\\ \hline
$[5,4;0,0,0]$ & $(1,0,5,0,0,0,0,0)$  & $(1,0,6,0,3,0,0,0)$ & $(1,0,6,0,4,0,1,0)$
\\ \hline
$[5,0;4,0,0]$ & $(0,0,3,0,4,0,0,0)$  & $(0,0,3,10,6,2,0,0)$ & $(0,0,4,11,10,2,0,0)$
\\ \hline
$[4,0;5,0,0]$ & $(0,0,0,0,6,0,0,0)$  & $(0,0,0,5,10,8,1,0)$ & $(0,0,0,6,19,14,2,0)$
\\ \hline
$[0,0;5,4,0]$ & $(0,0,0,0,0,0,1,0)$  & $(0,0,0,0,0,2,7,5)$ & $(0,0,0,0,0,11,44,40)$
\\ \hline
$[0,0;2,2,2]$ & $(0,0,4,0,1)$   & $(0,0,7,11,5)$  & $(0,0,8,19,16)$ 
\\ \hline
$[0,0;3,3,1]$  & $(0,0,1,0,3,0)$   & $(0,0,2,9,13,5)$  & $(0,0,2,14,35,23)$ 
\\ \hline
$[2,2;1,1,0]$ & $(3,0,13,0,0)$  & $(3,3,22,7,1)$ & $(3,3,23,13,5)$
\\ \hline 
$[1,1;2,2,0]$ &  $(0,0,12,0,1)$ & $(0,1,19,20,6)$  & $(0,1,22,33,20)$
\\ \hline 
$[3,2;1,1,0]$ & $(3,0,26,0,0,0)$ & $(3,3,35,17,7,0)$ & $(3,3,36,23,16,3)$
\\ \hline 
$[1,1;3,2,0]$ & $(0,0,9,0,6,0)$    & $(0,0,12,33,30,8)$  & $(0,0,13,46,78,40)$ 
\\ \hline  
$[1,1;1,1,1]$ & $(4,0,10,0)$ & $(4,6,20,5)$  & $(4,6,24,11)$
\\ \hline
$[2,1;1,1,1]$ & $(4,0,28,0,0)$ & $(4,6,44,21,3)$  & $(4,6,48,35,13)$
\\ \hline
$[1,1;2,1,1]$ & $(1,0,27,0,1)$ & $(1,4,41,34,8)$ & $(1,4,47,55,28)$
\\ \hline
$[1,1;3,1,1]$ & $(0,0,26,0,10,0)$ & $(0,1,32,71,52,11)$ & $(0,1,36,95,131,58)$
\\ \hline 
\end{tabular}
}

\bigskip

\centerline{\begin{tabular}{ |x{2.0cm}|x{4cm}|x{4.7cm}|}\hline
charges & $N=2$  & $N=\infty$
\\ \hline 
$[6,5;0,0,0]$  & $(1,0,9,0,1,0,0,0,0,0)$  & $(1,0,10,0,12,0,4,0,1,0)$
\\ \hline 
$[3,3;3,0,0]$  & $(0,0,23,0,19,0,0,0)$  & $(0,1,26,53,90,65,28,8)$
\\ \hline 
$[3,0;3,3,0]$  & $(0,0,3,0,21,0,1,0)$  & $(0,0,4,30,121,158,83,11)$
\\ \hline 
$[0,0;3,3,2]$  & $(0,0,1,0,4,0,1)$  & $(0,0,1,11,60,104,61)$
\\ \hline 
$[0,0;3,3,3]$  & $(0,0,0,0,4,0,3,0)$  & $(0,0,0,3,49,175,258,131)$
\\ \hline
$[1,1;2,2,1]$  & $(0,0,41,0,10,0)$  & $(0,1,58,128,170,72)$
\\ \hline 
$[2,1;2,1,1]$   & $(1,0,58,0,5,0)$ & $(1,4,81,118,114,35)$
\\ \hline 
$[2,2;1,1,1]$   & $(4,0,67,0,1,0)$  & $(4,6,94,92,77,20)$
\\ \hline 
$[2,2;2,1,1]$ & $(1,0,114,0,28,0,0)$ & $(1,4,140,242,382,237,60)$
\\ \hline 
$[2,1;2,2,1]$& $(0,0,77,0,42,0,0)$ & $(0,1,95,236,465,352,100)$ 
\\ \hline 
$[1,1;2,2,2]$& $(0,0,46,0,43,0,1)$ & $(0,0,54,191,508,515,199)$
\\ \hline 
\end{tabular}
}

\end{document}